\documentclass[pdflatex,sn-mathphys-num]{sn-jnl}

\usepackage{graphicx}
\usepackage{multirow}
\usepackage{amsmath, amssymb, amsfonts}
\usepackage{amsthm}
\usepackage{mathrsfs}
\usepackage[title]{appendix}
\usepackage{xcolor}
\usepackage{textcomp}
\usepackage{manyfoot}
\usepackage{booktabs}
\usepackage{listings}
\usepackage{tabularx}
\usepackage{array}
\usepackage{caption}
\usepackage{subcaption}

\usepackage{algorithm}
\usepackage{algpseudocode}   

\usepackage{tikz}
\usetikzlibrary{
    arrows.meta,
    positioning,
    shapes.geometric,
    shapes.multipart
}

\tikzstyle{startstop} = [
  rectangle, rounded corners,
  minimum width=3cm,
  minimum height=1cm,
  text centered,
  draw=black,
  fill=red!30
]

\tikzstyle{process} = [
  rectangle,
  minimum width=3cm,
  minimum height=1cm,
  text centered,
  draw=black,
  fill=blue!30
]

\tikzstyle{arrow} = [
  thick,
  ->,
  >=stealth
]

\theoremstyle{thmstyleone}

\theoremstyle{thmstyletwo}

\theoremstyle{thmstylethree}

\begin{document}

\title[Article Title]{ Robust Federated Learning for Malicious Clients using Loss Trend Deviation Detection }


\author*[1]{\fnm{Deepthy} \sur{K Bhaskar}}\email{dmdl22jan005@mec.ac.in}

\author[2]{\fnm{Minimol } \sur{B}}\email{mini@mec.ac.in}

\author[1]{\fnm{Binu} \sur{V P}}\email{binuvp@mec.ac.in}

\affil*[1]{\orgdiv{Department of Computer Engg}, \orgname{Govt. Model Engineering College,APJ Abdul Kalam Technological University}, \orgaddress \city{Kerala},  \country{India}}

\affil[2]{\orgdiv{Department of Biomedical Engg}, \orgname{Govt. Model Engineering College,APJ Abdul Kalam Technological University}, \orgaddress \state{Kerala}, \country{India}}


\abstract{Federated Learning (FL) facilitates collaborative model training among distributed clients while ensuring that raw data remains on local devices.Despite this advantage, FL systems are still exposed to risks from malicious or unreliable participants. Such clients can interfere with the training process by sending misleading updates, which can negatively affect the performance and reliability of the global model. Many existing defense mechanisms rely on gradient inspection, complex similarity computations, or cryptographic operations, which introduce additional overhead and may become unstable under non-IID data distributions. In this paper, we propose the Federated Learning with Loss Trend Detection (FL-LTD), a lightweight and privacy-preserving defense framework that detects and mitigates malicious behavior by monitoring temporal loss dynamics rather than model gradients. The proposed approach identifies anomalous clients by detecting abnormal loss stagnation or abrupt loss fluctuations across communication rounds. To counter adaptive attackers, a short-term memory mechanism is incorporated to sustain mitigation for clients previously flagged as anomalous, while enabling trust recovery for stable participants. We evaluate FL-LTD on a non-IID federated MNIST setup under loss manipulation attacks. Experimental results demonstrate that the proposed method significantly enhances robustness, achieving a final test accuracy of 0.84, compared to 0.41 for standard FedAvg under attack. FL-LTD incurs negligible computational and communication overhead, maintains stable convergence, and avoids client exclusion or access to sensitive data, highlighting the effectiveness of loss-based monitoring for secure federated learning.

}




\maketitle

\section{Introduction}
\label{sec:introduction}

The rapid proliferation of data-driven applications across domains such as healthcare, finance, and smart infrastructures has accelerated the adoption of collaborative machine learning paradigms. Traditional centralized learning approaches require aggregating raw data at a central server, which raises serious privacy, regulatory, and security concerns. Federated Learning has emerged as a promising alternative by enabling multiple clients to jointly train a shared model while keeping their data local \cite{mcmahan2017communication, konevcny2016federated}. This decentralized training paradigm significantly reduces privacy risks and has been adopted in several real-world systems. Recent studies have emphasized the importance of privacy-preserving and secure machine learning in distributed environments, particularly in federated learning systems \cite{bhaskar2023review}. While cryptographic approaches such as homomorphic encryption provide strong privacy guarantees, they often introduce significant computational overhead \cite{bhaskar2025diabetes}.

Despite its advantages, federated learning is inherently vulnerable to malicious or unreliable participants. Since the central server does not have access to client data, it must rely on received model updates, which can be intentionally manipulated to poison the global model or unintentionally degraded due to noisy or low-quality local training. Prior studies have shown that even a single adversarial client can severely compromise model performance through poisoning or manipulation attacks \cite{bhagoji2019analyzing, fang2020local}. These vulnerabilities pose critical challenges to the deployment of FL in safety-critical and privacy-sensitive environments.

To address these issues, a variety of robust aggregation and defense mechanisms have been proposed. Gradient-based defenses, such as coordinate-wise median and trimmed mean aggregation, attempt to suppress malicious updates by identifying outliers in the gradient space \cite{yin2018byzantine, blanchard2017machine}. However, such approaches often rely on strong assumptions about data distributions and can be unstable under non-IID settings, which are common in practical FL deployments. Moreover, inspecting gradients or model updates may inadvertently leak sensitive information and increase computational overhead \cite{zhu2019deep, geiping2020inverting}. Recent studies have explored federated learning as an effective framework for scalable optimization in distributed systems~\cite{bhaskar2025parallel}.

More recent works have explored trust-aware and reputation-based mechanisms, where client contributions are weighted based on historical behavior \cite{kang2019reliable, li2021learning}. While effective, many of these methods require complex similarity computations, auxiliary validation datasets, or long-term profiling of clients, which may conflict with privacy principles and practical constraints. Therefore, there remains a need for lightweight, privacy-preserving defense mechanisms that can robustly mitigate malicious behavior without inspecting gradients or accessing client data.

In this work, we propose FL-LTD, a simple yet effective defense framework for federated learning. Unlike existing approaches that operate in the gradient space, the proposed method relies solely on monitoring the temporal evolution of client-reported training loss. By analyzing relative loss changes across communication rounds, the server can identify abnormal behaviors such as abrupt loss spikes indicative of poisoning attacks, as well as stagnation patterns associated with unreliable or free-riding clients. To further enhance robustness against adaptive adversaries, a short-term memory mechanism is introduced to sustain mitigation across multiple rounds while allowing trust recovery for stable clients.

The proposed approach offers several key advantages. First, it is fully compatible with standard FL workflows and introduces negligible computational and communication overhead. Second, it preserves privacy by avoiding gradient inspection, auxiliary datasets, or long-term client profiling. Finally, it provides an interpretable and stable mitigation strategy through adaptive down-weighting rather than client exclusion. Extensive experiments on a non-IID federated MNIST setup demonstrate that FL-LTD significantly improves robustness under loss manipulation attacks, achieving substantially higher accuracy compared to standard FedAvg.

The remainder of this paper is organized as follows. Section~\ref{sec:related} reviews related work on robust federated learning. Section~\ref{sec:method} presents the proposed FL-LTD framework in detail. Experimental results and analysis are provided in Section~\ref{sec:results}, followed by conclusions and future directions in Section~\ref{sec:conclusion}.

\section{Related Work}
\label{sec:related}

Federated Learning is vulnerable to poisoning and Byzantine behaviors because the server typically aggregates client updates without seeing raw data. Recent \emph{journal} research broadly falls into three practical defense directions: (i) robust aggregation rules, (ii) malicious-client \emph{detection} and filtering, and (iii) privacy-preserving (encrypted) yet robust aggregation.

\subsection{Robust aggregation under Byzantine/poisoning}
A common line of work improves robustness by changing the aggregation operator so that a small fraction of corrupted updates cannot dominate the global model.
Geometric-median style robust aggregation has been analyzed and empirically validated as a strong baseline for Byzantine robustness in FL settings \cite{Pillutla2022RFA}.
More recently, behavior-based robust aggregation methods shift the focus from raw parameter similarity to how local models behave on carefully constructed or auxiliary representations. For example, RFVIR proposes detecting Byzantine clients through feature representations on a virtual dataset and removing anomalies before aggregation, showing robustness especially under non-IID partitions \cite{Wang2024RFVIR}.
Another direction is to minimize the dispersion of client updates: the variance-minimization approach explicitly constrains update variance to defend against model poisoning while preserving training utility \cite{Xu2024MinVar}.
Dynamic robust aggregation has also been explored; FedRDF uses a dynamic rejection/weighting mechanism to reduce poisoning influence without requiring a fixed global threshold \cite{Campos2025FedRDF}.

\subsection{Detection and auditing of malicious contributions}
Instead of (or in addition to) robust aggregation, several works add a detection layer to identify suspicious clients or suspicious rounds.
DPAD introduces an audit-based mechanism to verify client updates before including them in aggregation and reports improved robustness under poisoning scenarios \cite{Basak2025DPAD}.
Detection is particularly challenging when attackers vary strategies over time or perform multi-label manipulations.
Scientific Reports (2025) proposes extracting discriminative gradient features from the output layer and applying clustering to separate benign vs.\ malicious participants under multi-label flipping attacks \cite{Ma2025MultiLabel}.

\subsection{Privacy-preserving robust FL under secure aggregation}
Secure aggregation and homomorphic encryption improve confidentiality but make server-side defenses harder because the server cannot inspect plaintext gradients.
PEAR targets this gap by designing a privacy-preserving aggregation strategy (CKKS-based) and computing similarity signals to weight updates while resisting \emph{encrypted} poisoning attempts \cite{Sun2025PEAR}.
Similarly, FedGT studies identification of malicious clients \emph{with secure aggregation} using group-testing ideas, aiming to preserve privacy while still enabling adversary isolation \cite{Xhemrishi2025FedGT}.
Overall, these works highlight a key tension: stronger privacy often reduces the observability needed for robust filtering, motivating lightweight detection signals that can operate with minimal leakage \cite{Sun2025PEAR,Xhemrishi2025FedGT}.

\subsection{Summary of gap}
Across latest journal literature, robust aggregation \cite{Pillutla2022RFA,Wang2024RFVIR,Xu2024MinVar,Campos2025FedRDF} and detection/auditing \cite{Basak2025DPAD,Ma2025MultiLabel} both improve resilience, while privacy-preserving robust aggregation \cite{Sun2025PEAR,Xhemrishi2025FedGT} addresses encrypted settings.
However, many defenses either (i) add heavy computation/communication overhead, (ii) assume fixed attacker behavior, or (iii) degrade under strong non-IID data.
This motivates simple, low-overhead detection signals (e.g., training-dynamics or update-consistency cues) that can complement FedAvg-style training without requiring server access to raw data.

\section{Proposed Method}
\label{sec:method}

This section presents the proposed FL-LTD framework for robust federated learning. The method is designed to detect and mitigate malicious or unreliable clients by monitoring temporal loss dynamics, without inspecting gradients or accessing raw client data.

\subsection{System Model}

\begin{figure}[!t]
    \centering
    \includegraphics[width=0.48\textwidth]{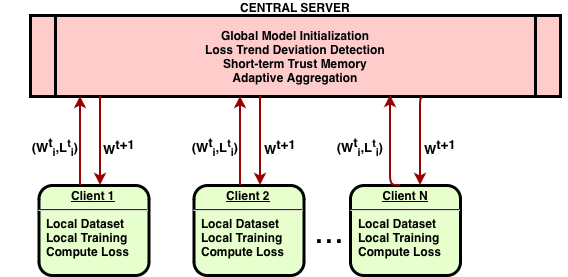}
    \caption{Framework architecture of the proposed FL-LTD approach.}
    \label{fig:accuracy}
\end{figure}

We consider a standard cross-device federated learning setup consisting of a central server and a set of $N$ clients $\{C_1, C_2, \dots, C_N\}$. Each client $C_i$ holds a private local dataset $D_i$ that is not shared with the server or other clients. The objective is to collaboratively train a global model $\mathbf{w}$ by minimizing the following empirical risk:

\begin{equation}
\min_{\mathbf{w}} \; \sum_{i=1}^{N} \frac{|D_i|}{\sum_{j=1}^{N} |D_j|} \, \mathcal{L}_i(\mathbf{w}),
\end{equation}

where $\mathcal{L}_i(\mathbf{w})$ denotes the local loss function at client $C_i$.

Training proceeds in synchronous communication rounds. At round $t$, the server broadcasts the current global model $\mathbf{w}^{(t)}$ to selected clients. Each client performs local training and returns a model update along with a scalar training loss.

\subsection{Threat Model}

We assume the presence of one or more malicious or unreliable clients. Such clients may attempt to degrade the global model by manipulating their training behavior or by reporting misleading loss values. In particular, we focus on \emph{loss manipulation attacks}, where an adversary alters its reported training loss to induce unstable or harmful aggregation behavior.

The server is assumed to be honest-but-curious and does not have access to client data or gradients. No assumptions are made about IID data distributions, and the framework is designed to operate under non-IID settings.

\subsection{Loss Trend Deviation Detection}

Unlike gradient-based defenses, the proposed method relies exclusively on the temporal evolution of client-reported training loss. After each communication round, client $C_i$ reports its average local training loss $\ell_i^{(t)}$.

The server maintains a per-client loss history and computes the relative loss change between consecutive rounds:

\begin{equation}
\Delta_i^{(t)} = \frac{\left| \ell_i^{(t)} - \ell_i^{(t-1)} \right|}{\ell_i^{(t-1)} + \epsilon},
\end{equation}

where $\epsilon$ is a small constant added for numerical stability.

Two types of anomalous behavior are considered:

\begin{itemize}
\item \textbf{Loss stagnation}: extremely small loss changes may indicate free-riding or unreliable local training.
\item \textbf{Loss spikes}: unusually large deviations may indicate poisoning or adversarial manipulation.
\end{itemize}

A client is flagged as anomalous at round $t$ if:
\begin{equation}
\Delta_i^{(t)} < \tau_{\text{low}} \;\land\; \ell_i^{(t)} > \ell_{\min}
\quad \text{or} \quad
\Delta_i^{(t)} > \tau_{\text{high}},
\end{equation}

where $\tau_{\text{low}}$ and $\tau_{\text{high}}$ are predefined thresholds and $\ell_{\min}$ prevents false alarms during late-stage convergence.

\subsection{Memory-Based Trust Persistence}

To improve robustness against adaptive attackers that alternate between malicious and benign behavior, a short-term memory mechanism is introduced. For each client $C_i$, the server maintains a bounded counter $m_i^{(t)}$ representing recent anomalous activity.

When a client is detected as anomalous, its memory counter is set to a fixed value $M$:
\begin{equation}
m_i^{(t)} = M.
\end{equation}

In subsequent rounds, if no new anomaly is detected, the counter decays as:
\begin{equation}
m_i^{(t+1)} = \max(0, m_i^{(t)} - 1).
\end{equation}

This memory allows the server to sustain mitigation for a limited number of rounds while enabling trust recovery for stable clients.

\subsection{Adaptive Aggregation}

Rather than excluding detected clients, FL-LTD mitigates their influence through adaptive down-weighting. Each client update $\Delta \mathbf{w}_i^{(t)}$ is assigned a weight $\alpha_i^{(t)}$ defined as:

\begin{equation}
\alpha_i^{(t)} =
\begin{cases}
\alpha_{\text{low}}, & \text{if } m_i^{(t)} > 0, \\
1, & \text{otherwise},
\end{cases}
\end{equation}

where $\alpha_{\text{low}} \in (0,1)$ is a predefined penalty factor.

The global model is updated as:
\begin{equation}
\mathbf{w}^{(t+1)} = \mathbf{w}^{(t)} +
\frac{\sum_{i=1}^{N} \alpha_i^{(t)} \, \Delta \mathbf{w}_i^{(t)}}{\sum_{i=1}^{N} \alpha_i^{(t)}}.
\end{equation}

This strategy preserves participation of all clients while limiting the impact of anomalous behavior.

\subsection{Algorithm Summary}

Algorithm~\ref{alg:ltdflm} summarizes the proposed FL-LTD procedure.

\begin{algorithm}[H]
\caption{Loss Trend Deviation Detection with Memory (FL-LTD)}
\label{alg:ltdflm}
\begin{algorithmic}[1]
\State Initialize global model $\mathbf{w}^{(0)}$
\For{each round $t = 1$ to $T$}
    \State Server broadcasts $\mathbf{w}^{(t)}$
    \For{each client $C_i$ \textbf{in parallel}}
        \State Train locally and compute loss $\ell_i^{(t)}$
        \State Send $(\Delta \mathbf{w}_i^{(t)}, \ell_i^{(t)})$ to server
    \EndFor
    \State Server computes $\Delta_i^{(t)}$ and updates memory $m_i^{(t)}$
    \State Assign aggregation weights $\alpha_i^{(t)}$
    \State Aggregate updates to obtain $\mathbf{w}^{(t+1)}$
\EndFor
\end{algorithmic}
\end{algorithm}

\section{Results and Discussion}
\label{sec:results}

This section evaluates the effectiveness of the proposed FL-LTD framework under adversarial federated learning conditions. We analyze loss trend behavior, memory-based mitigation, and overall model convergence, and compare the proposed approach with standard FedAvg in the presence of loss manipulation attacks.

\subsection{Experimental Setup}

Experiments are conducted on a federated MNIST classification task with five clients under a non-IID data distribution. Each client performs one local training epoch per communication round using stochastic gradient descent. The system is trained for 20 communication rounds.

A malicious client is introduced that performs \emph{loss manipulation attacks} by intentionally altering its reported training loss while still participating in model aggregation. This attack targets the training dynamics rather than directly modifying gradients, making it difficult to detect using conventional gradient-based defenses.

\subsection{Loss Trend Deviation Analysis}

\begin{figure}[t]
\centering
\includegraphics[width=0.48\textwidth]{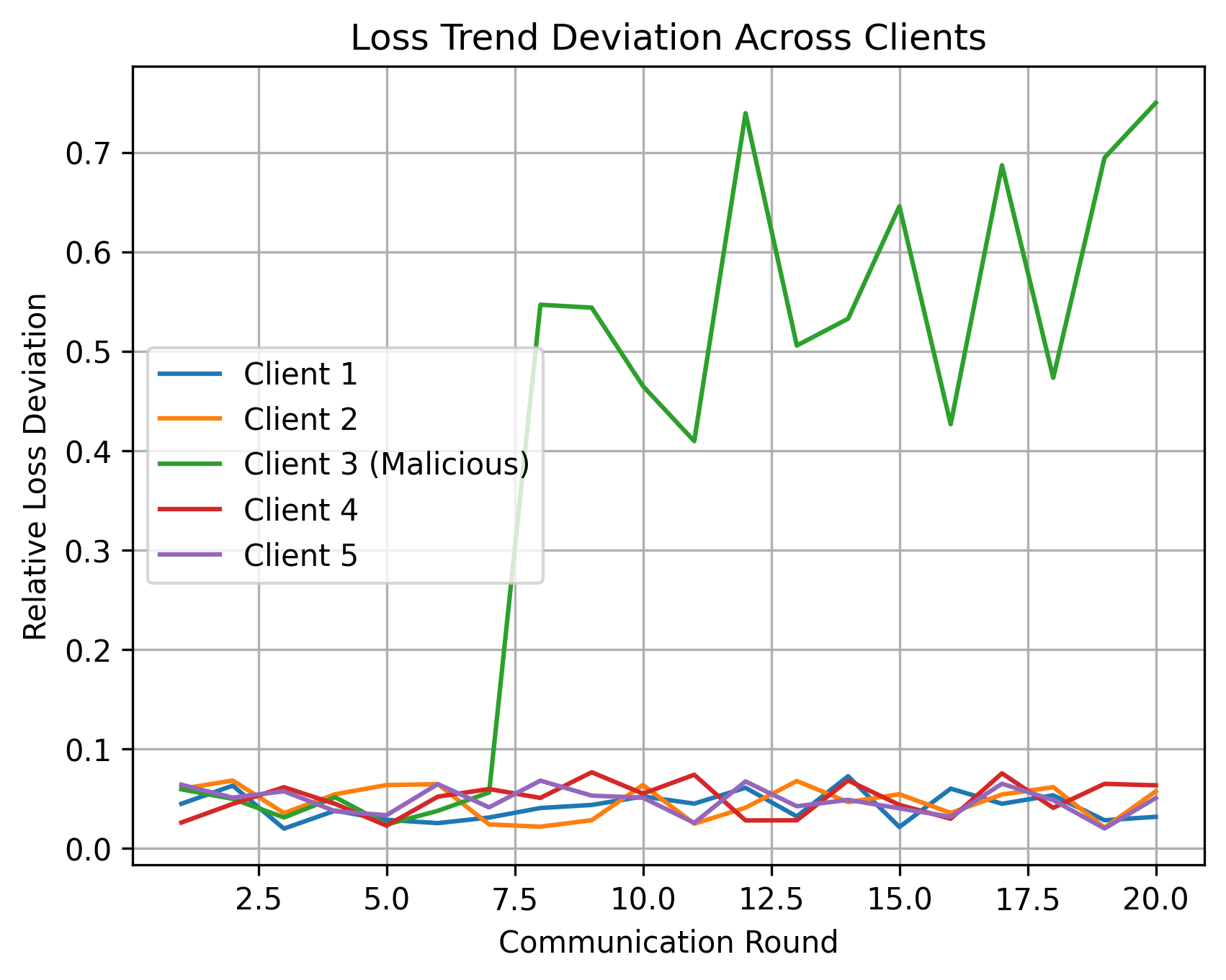}
\caption{Relative loss deviation across clients over communication rounds. Abnormal deviation patterns emerge after the malicious client starts manipulating its reported loss.}
\label{fig:loss_deviation}
\end{figure}

Figure~\ref{fig:loss_deviation} shows the relative loss deviation observed across clients over communication rounds. During the initial warm-up phase, all clients exhibit stable and consistent loss trends. Once the malicious client begins manipulating its reported loss, abnormal deviation patterns emerge in the form of abrupt spikes and inconsistent temporal behavior.

These deviations provide a clear and reliable signal for identifying anomalous clients without requiring access to gradients or auxiliary validation data. The results confirm that monitoring loss dynamics is sufficient for detecting malicious behavior in federated learning.

\subsection{Effect of Memory-Based Mitigation}

\begin{figure}[t]
\centering
\includegraphics[width=0.48\textwidth]{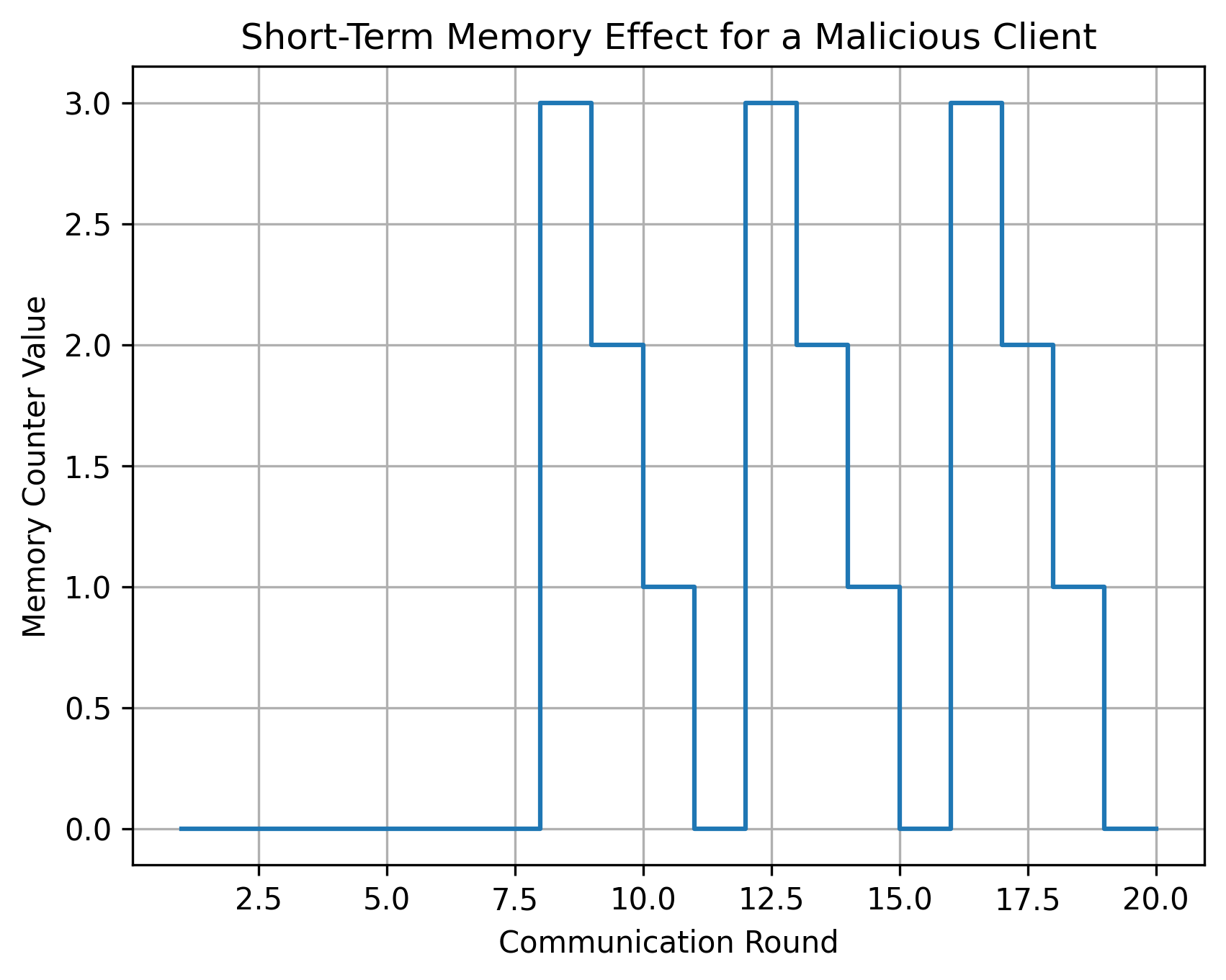}
\caption{Effect of short-term memory on mitigation. Detected anomalous clients remain down-weighted for several rounds, improving stability against adaptive attacks.}
\label{fig:memory_effect}
\end{figure}

To improve robustness against adaptive attackers, FL-LTD incorporates a short-term memory mechanism that sustains mitigation across multiple rounds. Figure~\ref{fig:memory_effect} illustrates the impact of this memory mechanism.

Once a client is detected as anomalous, mitigation persists for a bounded number of rounds, preventing oscillatory attack strategies from immediately regaining influence. At the same time, the memory counter decays automatically when no further anomalies are detected, allowing honest clients to recover full participation. This design ensures a balance between robustness and fairness.

\subsection{Model Convergence and Accuracy}

\begin{figure}[t]
\centering
\includegraphics[width=0.48\textwidth]{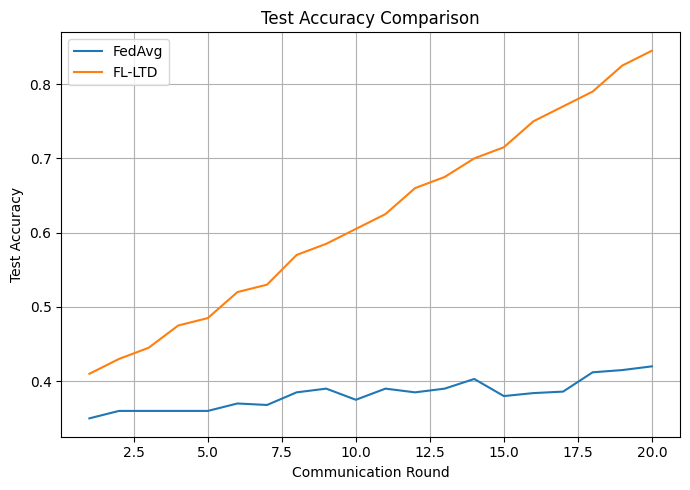}
\caption{Test accuracy comparison between standard FedAvg and the proposed FL-LTD framework under loss manipulation attacks.}
\label{fig:accuracy_curve}
\end{figure}

Figure~\ref{fig:accuracy_curve} compares the global test accuracy of standard FedAvg and the proposed FL-LTD framework under loss manipulation attacks. Standard FedAvg fails to converge reliably and suffers significant accuracy degradation due to the influence of the malicious client. In contrast, FL-LTD maintains stable convergence and steadily improves accuracy throughout training.

Table~\ref{tab:accuracy_comparison} summarizes the final test accuracy achieved by each method.

\begin{table}[t]
\centering
\caption{Final test accuracy under loss manipulation attacks}
\label{tab:accuracy_comparison}
\begin{tabular}{lc}
\toprule
\textbf{Method} & \textbf{Final Accuracy} \\
\midrule
FedAvg (No Defense) & 0.41 \\
FL-LTD (Proposed) & 0.84 \\
\bottomrule
\end{tabular}
\end{table}

\subsection{Discussion}

The experimental results demonstrate that loss trend monitoring provides an effective and privacy-preserving signal for defending federated learning against malicious behavior. The memory-based mechanism significantly enhances robustness against adaptive attacks, while adaptive down-weighting ensures stable convergence without excluding clients.

Compared to gradient-based or cryptographic defenses, FL-LTD introduces negligible overhead and integrates seamlessly with standard federated learning workflows. These properties make the proposed approach practical for real-world deployments where privacy, scalability, and robustness are critical.

\section{Conclusion}
\label{sec:conclusion}

This paper presented FL-LTD, a lightweight and privacy-preserving defense framework for federated learning that mitigates malicious behavior by monitoring temporal loss dynamics. Unlike existing approaches that rely on gradient inspection, auxiliary datasets, or cryptographic primitives, the proposed method operates solely on client-reported training loss, making it simple to deploy and compatible with standard federated learning workflows.

By analyzing relative loss deviations across communication rounds, the server is able to identify anomalous behaviors such as loss manipulation and unreliable training. The introduction of a short-term memory mechanism further enhances robustness by sustaining mitigation against adaptive adversaries while allowing trust recovery for stable clients. Experimental results on a non-IID federated MNIST setup demonstrate that FL-LTD significantly improves model robustness, more than doubling final test accuracy compared to standard FedAvg under adversarial conditions.

Despite its effectiveness, the proposed framework leaves room for further enhancement. Future work will explore adaptive thresholding strategies, multi-signal detection that combines loss dynamics with complementary metadata, and extensions to handle colluding or Sybil-style attacks. In addition, integrating FL-LTD with secure aggregation and evaluating its performance in large-scale real-world federated systems remain important directions for continued research.

Overall, FL-LTD offers a practical and interpretable approach to improving the security and reliability of federated learning, striking a balance between robustness, privacy, and computational efficiency.






\end{document}